\documentclass[fleqn,10pt,twocolumn]{SICE_FES25}
\makeatletter
\def\caption@documentclass{elsarticle} % subcaption
\makeatother

\usepackage{textcomp} % ~
\newcommand{\textapprox}{\raisebox{0.5ex}{\texttildelow}} % ~
\usepackage{amsmath}
\usepackage{subcaption}
\usepackage{caption}
\usepackage{overpic}
\usepackage{multirow}
\usepackage{booktabs} % 表 \toprule, \midrule, \bottomrule
\usepackage{siunitx} % degreeCelsius, SI
\usepackage{cleveref}
\crefname{equation}{Eq.}{Eqs.} % eqnarrayではなくequationを使用
\crefname{figure}{Fig.}{Figs.} % {環境名}{単数形}{複数形}
\crefname{table}{Tab.}{Tabs.}
\title{
Synchro-Thermography: Monitoring \textapprox 10 mK Facial Temperature Changes with Heartbeat Referencing for Physiological Sensing
}

\author{Nanami Kotani${}^{1\dagger*}$, Kuniharu Sakurada${}^{2*}$, Jiayi Xu${}^{3*}$, Masahiko Inami${}^{4*}$ and Yasuaki Monnai${}^{5*}$}
\speaker{Nanami Kotani}

\affils{${}^{*}$Research Center for Advanced Science and Technology, The University of Tokyo\\
${}^{1}$(E-mail: nanami.kotani@star.rcast.u-tokyo.ac.jp)\\
${}^{2}$(E-mail: kuniharu.sakurada@star.rcast.u-tokyo.ac.jp)\\
${}^{2}$(E-mail: xu.jiayi@star.rcast.u-tokyo.ac.jp)\\
${}^{3}$(E-mail: inami@star.rcast.u-tokyo.ac.jp)\\
${}^{4}$(E-mail: monnai@star.rcast.u-tokyo.ac.jp)\\
}
\abstract{% 目安 150-250 words
Infrared thermography has gained interest as a tool for non-contact measurement of blood circulation and skin blood flow due to cardiac activity. Particularly, blood vessels on the surface, such as on the back of the hand, are suited for visualization. However, standardized methodologies have not yet been established for areas such as the face and neck, where many blood vessels lie deeper beneath the surface, and external stimulation for measurement could be harmful. Here, we propose Synchro-Thermography for stable monitoring of facial temperature changes associated with heart rate variability. We conducted experiments with eight subjects and measured minute temperature changes with an amplitude of about \SI{10}{mK} on the forehead and chin. The proposed method improves the temperature resolution by a factor of 2 or more, and can stably measure skin temperature changes caused by blood flow. This skin temperature change could be applied to physiological sensing, such as blood flow changes due to injury or disease, or as an indicator of stress.
}


\begin{document}

\maketitle

%-----------------------------------------------------------------------

\section{Introduction}
Infrared thermography is gaining interest as a tool for non-contact measurement of blood circulation and skin blood flow due to cardiac activity. While typical thermometers provide one-dimensional data, thermal images measured with infrared thermography provide two-dimensional data, and thermal videos offer time-series sequences of two-dimensional data, which can be regarded as three-dimensional data consisting of spatial (x, y) and temporal (t) dimensions. Combining thermal videos with other anatomical and functional information has the potential for accurate diagnosis and reliable treatment in medical applications \cite{Chen2019BEL,Jose2022FP,Lahiri2012IPT}. Temperature measurement using thermography is highly practical because of its non-contact nature, the capability of measuring a large number of people simultaneously in a hygienic manner, and the simplicity of the equipment setup. However, measuring minute changes in body temperature with infrared thermography is difficult not only because of noisy artifacts during measurement but also environmental changes and individual factors like body motions \cite{Fernandez2015IPT}. Especially in areas such as the face and neck, it is difficult to measure because many blood vessels lie deeper under the surface.
In addition, external stimuli to enhance contrast, as in previous studies \cite{Bouzida2009JTB,Liu2012TF,Sagaidachnyi2017PM}, can be uncomfortable and harmful. In this study, we propose Synchro-Thermography, which utilizes the heartbeat as a reference signal to enhance the signal-to-noise ratio of the measurement. The proposed method can measure minute changes in body temperature associated with fluctuations in heart rate. Skin temperature changes measured by the proposed method could be applied to physiological and stress sensing involving blood flow changes.

\section{Related work}
The relation between body temperature and physiological states has been discussed.
The human body temperature varies depending on environmental factors such as ambient temperature and relative humidity, as well as factors due to individual characteristics \cite{Fernandez2015IPT}. The individual factors can be classified into intrinsic and extrinsic factors. The intrinsic factors refer to basic characteristics and endogenous factors such as gender, disease, skin blood flow, and emotions. Extrinsic factors are exogenous factors that temporarily affect the body, such as alcohol, caffeine, food, water intake, and exercise. Thus, the body temperature is one of the fundamental physiological indicators.
It is well known that heart rate and body temperature are positively correlated due to thermal homeostasis \cite{Charkoudian2003MCP,Chen2019BEL}. Thus, since skin temperature is particularly influenced by blood flow, many studies have been conducted to measure blood flow and visualize blood vessels using thermography.

In the extremities, there are many blood vessels that are large enough to be visible on the skin surface, and it is easy to apply external stimuli such as pressure or cold stress. For this reason, many studies have been conducted to visualize blood vessels using thermographies \cite{Bouzida2009JTB,Lahiri2012IPT,Liu2012TF,Sagaidachnyi2017PM}.
The clinical applications of thermographic monitoring of blood circulation can be divided into inflammation-based (increased blood flow) and perfusion-based (decreased blood flow) approaches. Inflammation-based scenarios include hidradenitis suppurativa and lupus erythematosus, while perfusion-based scenarios include ischemic diabetic foot ulcer and surgical free and rotational flaps \cite{Jose2022FP}.
The recording of vasomotor alterations through thermography has been applied not only in human medicine but also in veterinary medicine \cite{Alejandro2020JTB}. In human medicine, local anesthesia is believed to cause vasodilation, a temporary increase in blood flow, and an increase in skin temperature. Therefore, measurement of vasomotor alterations could be used in veterinary medicine to evaluate whether anesthesia was performed correctly. Veterinary medicine is also investigating the use of thermal imaging cameras to monitor tissue grafts in lesions caused by burns, malformations, or tumor surgery, where blood perfusion is essential.

However, most studies are limited to obtaining averaged thermal images. If it is possible to measure body temperature changes at each heartbeat as time series data, it will have a wider range of applications. For this reason, we propose Synchro-Thermography as a method to measure minute changes in body temperature associated with heart rate variability without external stimuli. We aim to construct a system that can measure changes in skin temperature, taking individual differences into account by significantly updating our previous study \cite{Kotani2024SICE}. We believe that the changes in skin temperature measured by the proposed method have potential applications in physiological information sensing, such as clinical applications in monitoring blood circulation.

\section{Proposed method}
\begin{figure}[t]
    \centering
    \includegraphics[width=\linewidth]{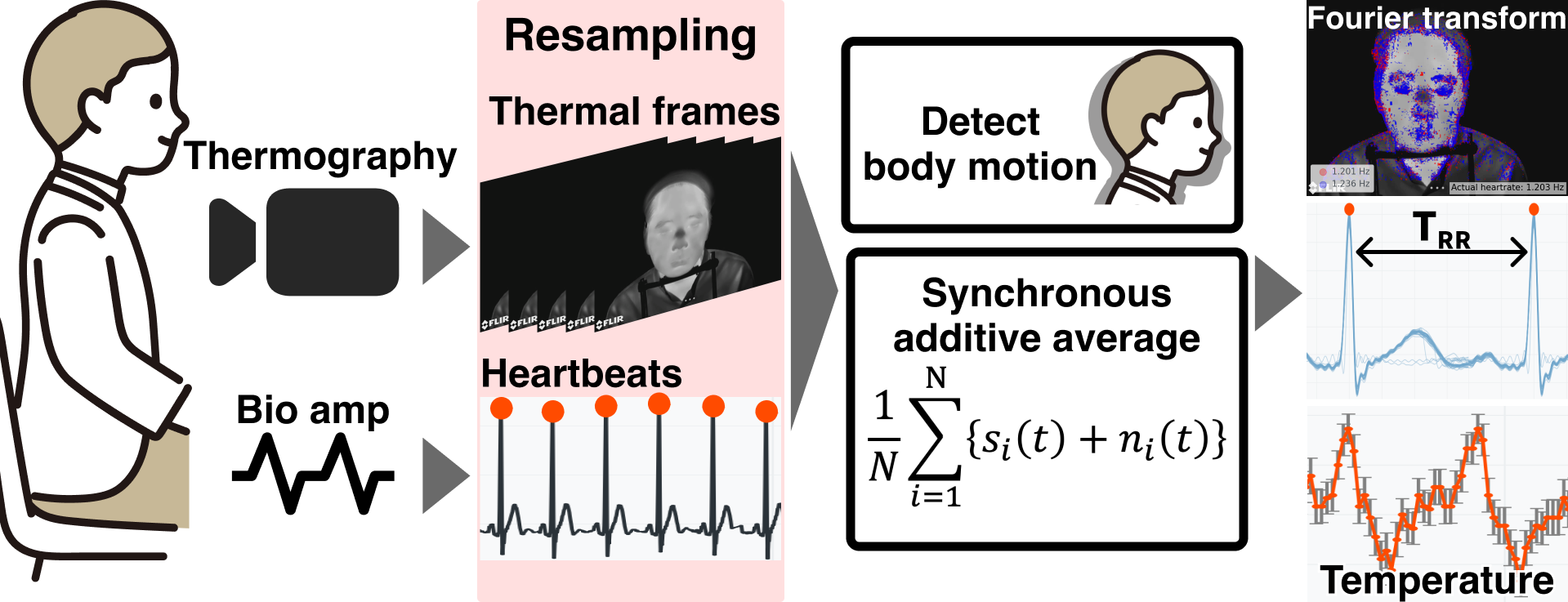}
    \caption{Overview of our proposed Synchro-Thermography using the heartbeats as a reference timing signal}
    \label{fig:overview}
\end{figure}

We propose Synchro-Thermography for measuring minute changes in body temperature associated with heart rate variability. An overview of the proposed method is illustrated in \cref{fig:overview}. Specifically, the following procedure is used to perform synchronous additive averaging of thermal images using the heartbeat as a timing signal.

\begin{enumerate}
    \item \textbf{Resampling:} Resample the heartbeats and thermal video into equally spaced time series data based on the measured mean heartbeat interval.
    \item \textbf{Body Movement Detection:} For a set of time-scaled thermal images, RIPOC \cite{Reddy1996TIP} is used to align the images and absorb the subject's body movement.
    \item \textbf{Averaging:} The sliding-window method is used to take a synchronous additive average over the aligned images.
    \item \textbf{Fourier transform:} Fourier transform the synchronous averaged thermal video in the time sequence direction to detect regions where the body temperature fluctuates at the same period as the heartbeat cycle.
\end{enumerate}

In this study, for the calculation of the heartbeat interval, the R-peaks were detected from the ECG based on Christov's algorithm \cite{Christov2004BEO}.
The principle of the proposed method is explained as follows. Firstly, we process the data using a sliding window method. Let $T_{win}$ be the window size and $T_{slide}$ be the slide size. Let $N_{slide}$ be the number of slides, and the total time $T_{all}$ of data used for analysis is the following \cref{eq:slidingWindow}.
\begin{equation}
    \begin{array}{rcl}
    T_{all} &=& T_{win} + N_{slide} \times T_{slide}
    \label{eq:slidingWindow}
    \end{array}
\end{equation}
Within the time $T_{win}$ to be analyzed, the number of heartbeats recorded is $N$ and the average RR interval during synchronous addition is $T_{RR}$. The average RR interval can be expressed as $T_{RR} = T_{win} / N$. In this case, the result of applying the proposed method with the sliding window method is thermal video data of time $T_{result}$, which can be expressed as \cref{eq:resultTime}.
\begin{equation}
    \begin{array}{rcl}
    T_{result} = T_{RR} \times (N_{slide}+1) \label{eq:resultTime}
    \end{array}
\end{equation}

Resampling and body motion detection are performed in batches for the total time $T_{all}$ of data used for analysis. Then, the data is divided into $(N_{slide}+1)$ time $T_{win}$ data, which is the number of slides $N_{slide}$ plus the first one from the time $T_{all}$. Perform synchronous additive averaging on the time $T_{win}$ data. The synchronous additive averaged data are concatenated into a single thermal video data of time $T_{result}$ and then Fourier transformed.

Next, we discuss synchronous additive averaging. The measured temperature $f_i(t)\mbox{ }(i=1,2,\ldots,N)$ resampled at the mean heartbeat period can be expressed as the sum of the true temperature $s_i(t)$ and the noise $n_i(t)$.
Therefore, the resulting synchronous additive average can be expressed as follows \cref{eq:synchronousAdditiveAverage}.
\begin{equation}
    \begin{array}{rcl}
    \bar{f}(t) = \frac{1}{N} \sum_{i=1}^N \{ s_i (t) + n_i (t) \} ~~ (0 \leq t < T_{RR}) \label{eq:synchronousAdditiveAverage}
    \end{array}
\end{equation}

The standard error $SEM$ was calculated for the synchronous additive average results based on \cref{eq:standardError}.
\begin{equation}
    \begin{array}{rcl}
    SEM &= \sqrt{\frac{1}{N^2} \sum^N_{i=1}(f_i - \bar{f})^2}  \label{eq:standardError}
    \end{array}
\end{equation}

\if0
The sliding window method is represented in the figure as shown in \cref{fig:slideWindow}.
\begin{figure}
    \centering
    \includegraphics[width=\linewidth]{fig/slideWindow.png}
    \caption{The sliding window method}
    \label{fig:slideWindow}
\end{figure}
\fi

To evaluate whether the proposed method can measure body temperature fluctuations associated with heart rate fluctuations, the cross-correlation between heart rate and body temperature was calculated. However, as shown in the following section (\cref{sec:results}), the cross-correlation was calculated using the cosine wave generated from the average heartbeat interval, taking into consideration that the waveform of the body temperature fluctuation is like a cosine wave, like the pulse rate.

\section{Experiment setup}
\subsection{Participants}
The experiment was conducted with eight participants. 
Before the experiment, we conducted a preliminary experiment with a healthy female in her 20s, who also participated in the experiment. 
Of these, five are male and three are female. All are in their 20s and in good health.
The series of experiments was conducted with the approval of the university's ethics review committee (approval number 23-466). Before participating in the experiment, the study was explained to the participants, and written consent for participation was obtained.

\subsection{Equipments and environments}
\begin{figure}[tb]
    \centering
    \begin{minipage}{.56\linewidth}
        \includegraphics[width=\linewidth]{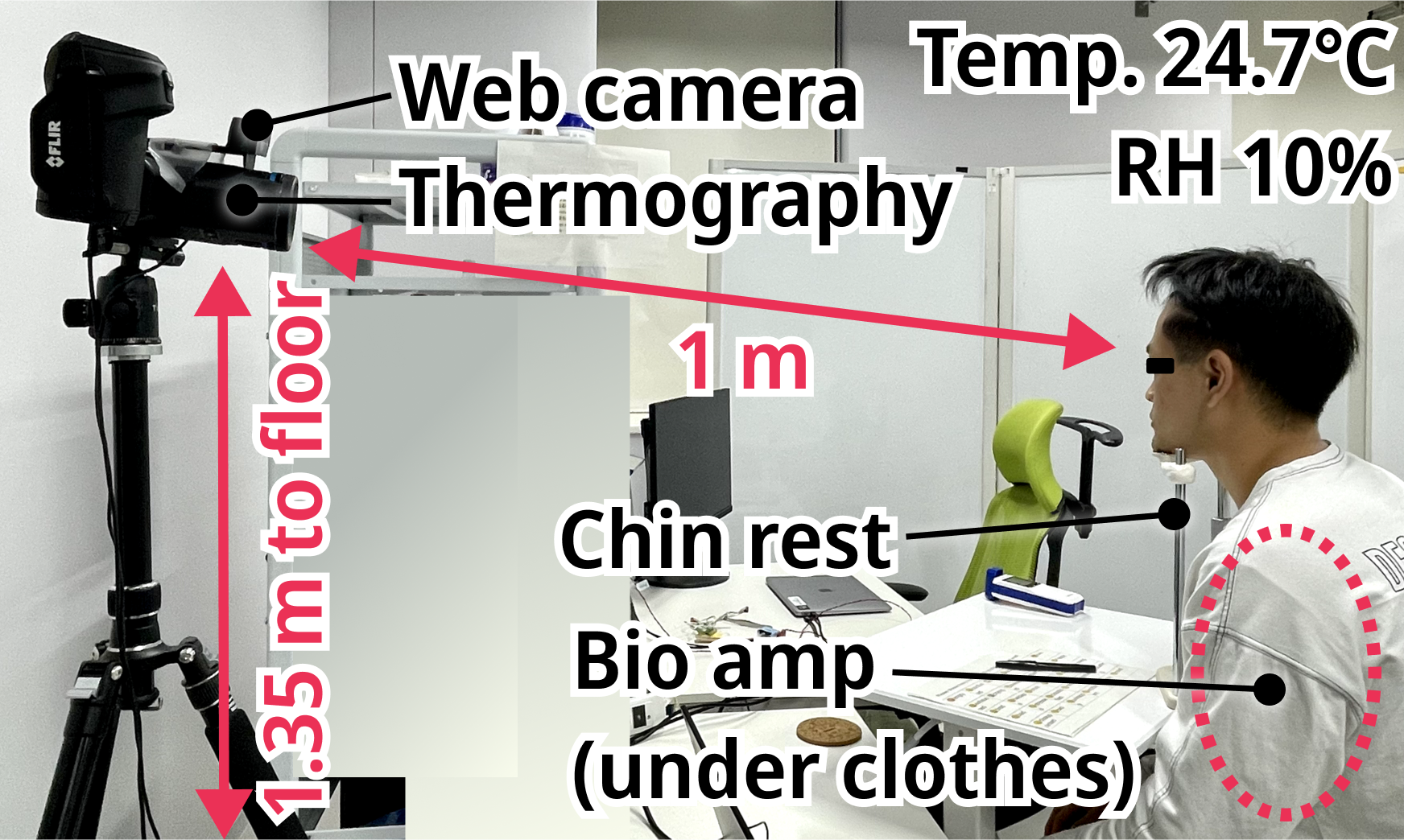}
        \subcaption{The experimental environment}
        \label{fig:experimentalEnv}
    \end{minipage}
    \begin{minipage}{.4\linewidth}
        \includegraphics[width=\linewidth]{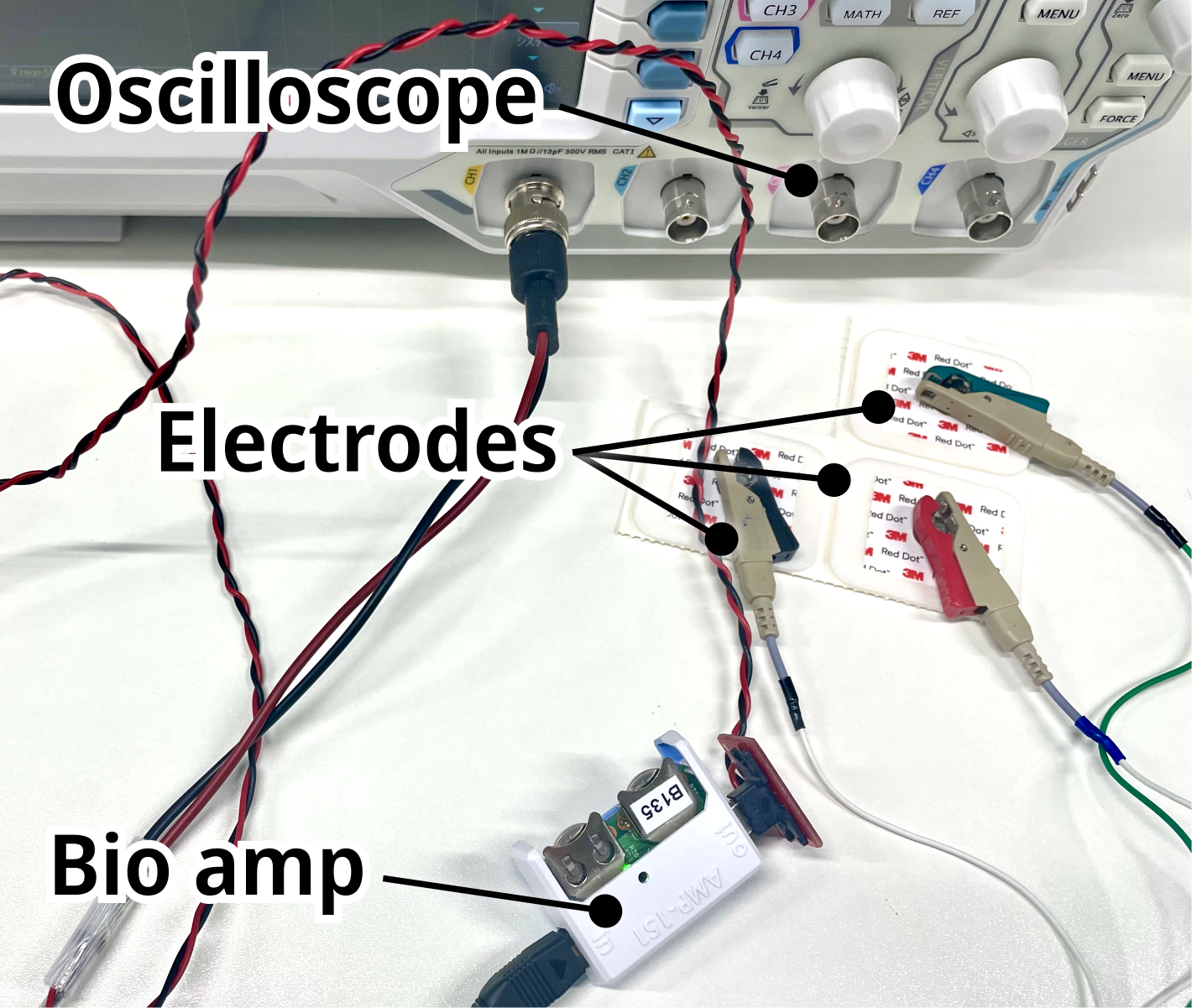}
        \subcaption{The biosignal amplifier}
        \label{fig:bioAmp}
    \end{minipage}
    \caption{The experimental environment and equipment}
\end{figure}

The experimental environment is shown in \cref{fig:experimentalEnv}. 
The experiment was conducted in a room with a mean room temperature of \SI{24.7}{\degreeCelsius} (SD \SI{0.32}{\degreeCelsius}) and a mean relative humidity of \SI{10.0}{\%} (SD \SI{0.0}{\%}). The experiments were performed with these measurement parameters properly set up on the thermography.
The room size was $\SI{6.2}{m} \times \SI{8.0}{m}$ and the curtains were closed to prevent direct sunlight. A partition was set up at \SI{2.7}{m} from the thermography to control the background environment.
The thermography was a FLIR T530 (FLIR Systems Inc.). A logicool C922n webcam (Logicool Co Ltd.) was attached to the top of the thermography to capture visible and thermal video simultaneously. Under the subject's clothing, an electrocardiogram is measured using a biosignal amplifier (AMP-151, ATR-Promotions Inc.) as shown in \cref{fig:bioAmp}. The biosignal amplifier is used by attaching three electrodes to the chest.
The thermography was placed at \SI{135}{cm} from the floor and \SI{100}{cm} from the subject. The display resolution of the thermography was $480 \times \SI{640}{pixels}$ and the frame rate was \SI{30}{fps}. The temperature resolution (noise equivalent temperature difference) is less than \SI{40}{mK} at \SI{30}{\degreeCelsius}.
The thermal video and ECG measurements are controlled by the same computer and time-synchronized based on the start time of imaging.

\subsection{Experimental procedure}
The following procedure was used in the experiment.
\begin{enumerate}
    \item \textbf{Informed consent:} Fill out the consent form and pre-participation questionnaire.
    \item \textbf{Preparation for the experiment:} Electrodes for heart rate measurement were attached, and the face was exposed. The patient is then allowed to rest for 10 minutes to acclimate to the room temperature.
    \item \textbf{Resting state measurement:} Heart rate and thermal video will be recorded for 3 minutes.
\end{enumerate}

The subjects are seated and place his/her head on the chin rest and fix his/her head in a comfortable position for the filming. Rest breaks were taken as needed between each record to relieve fatigue.

In a preliminary experiment, the images were taken from three directions: frontal face, right profile, and left profile. In the main experiment, the left profile was photographed based on the results of the preliminary experiment and a previous study that found no predominant temperature difference between the left and right sides of the face \cite{Haddad2016DR}.

\subsection{Parameter settings} \label{subsec:parameter}
\begin{table}[tb]
    \caption{Parameter settings}
    \label{tab:parameterSettings}
    \centering
    \begin{tabular}{ll} \toprule
        Parameter & Set value \\ \hline
        The total data size $T_{all}$& \SI{30}{\s} \\
        The window size $T_{win}$& 10 heartbeats \\
        The slide size $T_{slide}$& 3 heartbeats \\
        The number of slides $N_{slide}$& $(T_{all} - T_{win}) / T_{slide}$ \\
        \bottomrule
    \end{tabular}
\end{table}
The body temperature distribution was calculated from the average of one center pixel $(x, y)$ and 24 pixels around $(x-2, y-2)$ to $(x+2, y+2)$, for a total of 25 pixels.
In addition, data from 30 seconds of particularly low body motion out of a 3-minute session was used for analysis as $T_{all}$. Other parameters were set as in \cref{tab:parameterSettings}, based on the individual's heartbeat interval. $T_{win}, T_{slide}$, considering $T_{all}$, and the parameters were heuristically adjusted from the results of our previous study \cite{Kotani2024SICE} and our experimental data. These parameters should be adjusted to maximize the correlation between body temperature and heartbeats, taking into account factors such as the subject's age and experimental environment.

\section{Results and discussion} \label{sec:results}
\subsection{Preliminary experiments}
\begin{figure}[btp]
    \centering
    % \begin{subcaptiongroup}
    \begin{minipage}{.47\linewidth}
        \centering
        \subcaptionlistentry{Raw thermal image}
        \label{fig:rawFemale}
        \begin{overpic}[width=.7\linewidth]{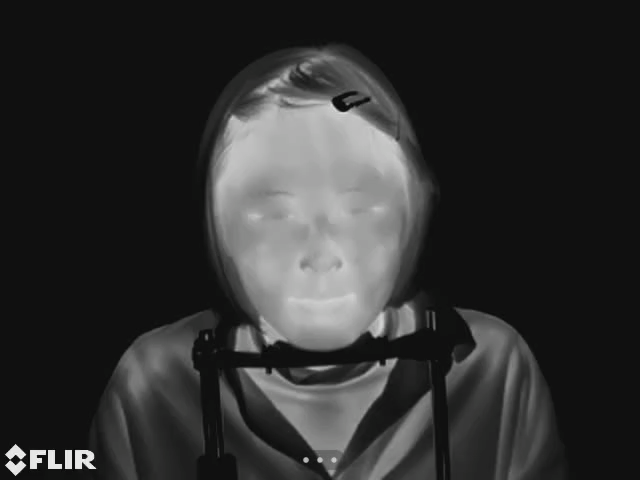}
            \put(-20,5){\colorbox{white}{\captiontext*{}}}
        \end{overpic}
    \end{minipage}
    \begin{minipage}{.47\linewidth}
        \centering
        \subcaptionlistentry{Proposed thermal image}
        \label{fig:synchroFemale}
        \begin{overpic}[width=.7\linewidth]{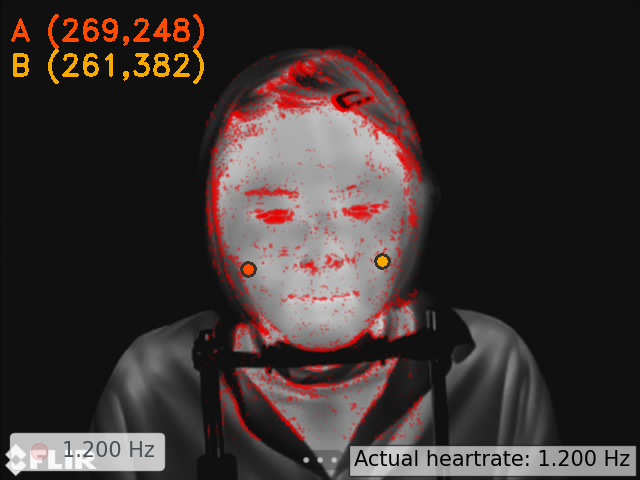}
            \put(-20,5){\colorbox{white}{\captiontext*{}}}
        \end{overpic}
    \end{minipage}
    \begin{minipage}{.47\linewidth}
        \subcaptionlistentry{Raw heartbeats}
        \label{fig:rawECGfemale}
        \begin{overpic}[width=\linewidth]{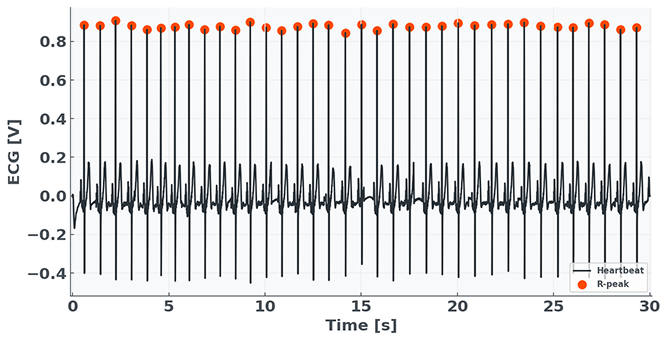}
            \put(0,5){\colorbox{white}{\captiontext*{}}}
        \end{overpic}
    \end{minipage}
    \begin{minipage}{.47\linewidth}
        \subcaptionlistentry{Normalized heartbeats}
        \label{fig:synchroECGfemale}
        \begin{overpic}[width=\linewidth]{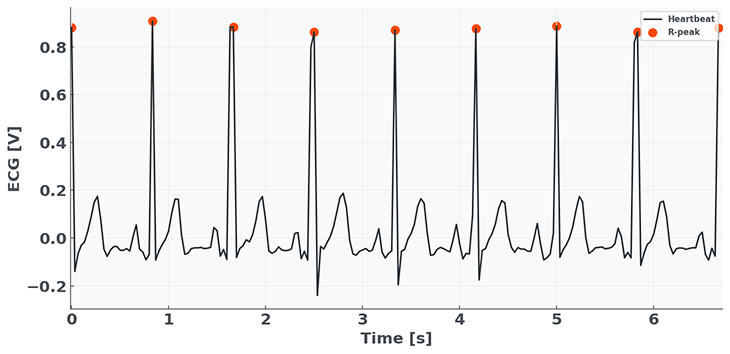}
            \put(0,5){\colorbox{white}{\captiontext*{}}}
        \end{overpic}
    \end{minipage}
    \begin{minipage}{.47\linewidth}
        \subcaptionlistentry{Raw temperature A}
        \label{fig:rawAfemale}
        \begin{overpic}[width=\linewidth]{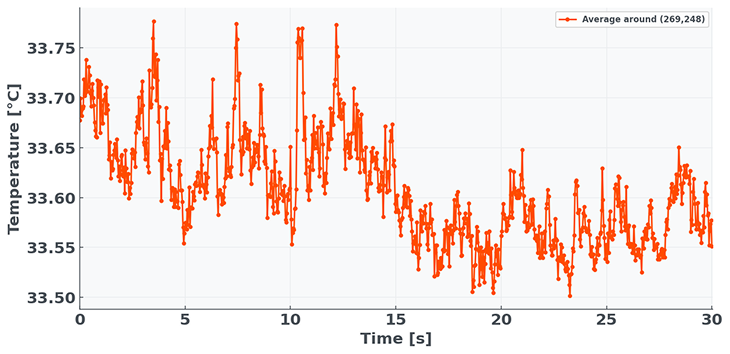}
            \put(0,5){\colorbox{white}{\captiontext*{}}}
        \end{overpic}
    \end{minipage}
    \begin{minipage}{.47\linewidth}
        \subcaptionlistentry{Proposed temperature A}
        \label{fig:synchroAfemale}
        \begin{overpic}[width=\linewidth]{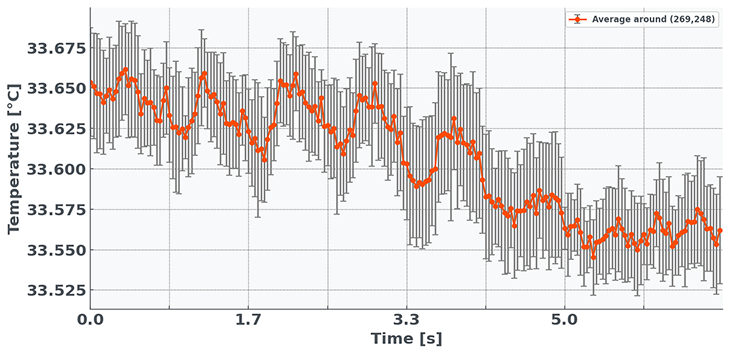}
            \put(0,5){\colorbox{white}{\captiontext*{}}}
        \end{overpic}
    \end{minipage}
    \begin{minipage}{.47\linewidth}
        \subcaptionlistentry{Raw temperature B}
        \label{fig:rawBfemale}
        \begin{overpic}[width=\linewidth]{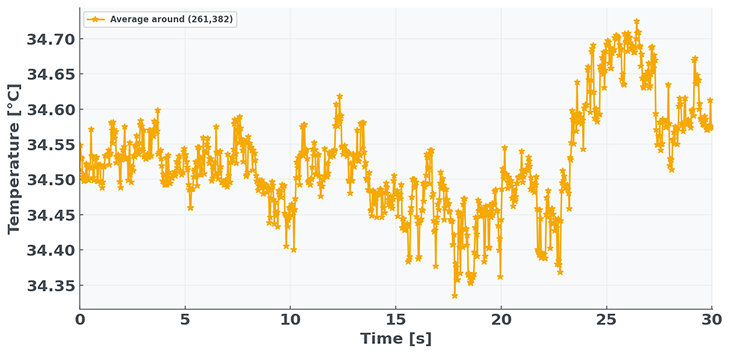}
            \put(0,5){\colorbox{white}{\captiontext*{}}}
        \end{overpic}
    \end{minipage}
    \begin{minipage}{.47\linewidth}
        \subcaptionlistentry{Proposed temperature B}
        \label{fig:synchroBfemale}
        \begin{overpic}[width=\linewidth]{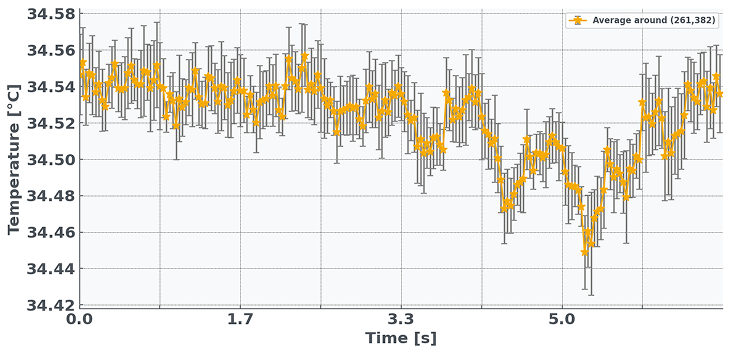}
            \put(0,5){\colorbox{white}{\captiontext*{}}}
        \end{overpic}
    \end{minipage}
    % \end{subcaptiongroup}
    \caption{Raw data in left column and results of the proposed method in right column: Female frontal face. \subref{fig:rawFemale}, \subref{fig:synchroFemale} The thermal video frame. \subref{fig:rawECGfemale} Heartbeats of raw data. \subref{fig:synchroECGfemale} Heartbeats resampled to the mean RR interval. \subref{fig:rawAfemale}, \subref{fig:synchroAfemale} Temperature at the right cheek A (269, 248). \subref{fig:rawBfemale}, \subref{fig:synchroBfemale} Temperature at the right cheek B (261, 382)} 
    \label{fig:preFemale}
\end{figure}
In preliminary experiments, frontal, right profile, and left profile images were taken, and the results for the frontal face are shown in \cref{fig:preFemale} as a representative example.
\cref{fig:synchroFemale} is the result of Fourier transforming the synchronous addition-averaged thermal video in the time-series column direction. The scatter plots are depicted in the areas that oscillate with the same period as the mean heartbeat period \SI{1.2}{Hz}, and the larger the amplitude, the denser the scatter plots are. As described in \cref{subsec:parameter}, the coordinates (pixel values) depicting the temperature graphs (\crefrange{fig:rawAfemale}{fig:synchroBfemale}) are plotted in \cref{fig:synchroFemale}.
The correlation coefficient between the heart rate variability shown in \cref{fig:synchroECGfemale} and the body temperature variability shown in \cref{fig:synchroAfemale,fig:synchroBfemale} is -0.5 for the body temperature of the right cheek A and -0.42 for the body temperature of the left cheek B. The correlation coefficient between the body temperature of the right cheek A and the heart rate is -0.5. Here, when the correlation coefficient is negative, it indicates that the heart rate and body temperature are in opposite phases.
\cref{fig:synchroAfemale,fig:synchroBfemale} can be inferred to be the body temperature fluctuation affected by the transverse facial artery. Here, the ECG has a clear steep high wave (R wave) as shown in \cref{fig:synchroECGfemale}. On the other hand, in \cref{fig:synchroAfemale,fig:synchroBfemale}, the influence of the R wave is more gradual, as in the pulse wave, and the waveform can be read as sinusoidal or sawtooth wave-like.

\subsection{Main experiments}
The correlation coefficients for each subject between temperatures and heartbeats after applying synchro-thermography are shown in \cref{tab:correlation}, summarizing the results of the preliminary experiment and the main experiment. We picked up two or three areas where the correlation coefficients became particularly large after the application of the proposed method. From \cref{tab:correlation}, it can be seen that the absolute value of the correlation coefficient increases after applying the proposed method, and the correlation between heart rate and body temperature becomes clearer. This result suggests that synchro-thermography could be able to extract the body temperature changes associated with the heart rate variability.

\begin{table}[tb]
    \caption{Correlation between temperatures and heartbeats}
    \label{tab:correlation}
    \centering
    \begin{tabular}{ll|cc} \toprule
        \multirow{2}{*}{Subject name}&Measurement&\multicolumn{2}{c}{Correlation} \\
        &area&Raw&Proposed \\ \hline
        Preliminary&Right cheek A&-0.13&-0.50 \\
        experiments&Left checck B&0.06&-0.42 \\ \hline
        \multirow{3}{*}{Subject 1}&Temple A&-0.11&-0.56 \\
        &Temple B&-0.06&0.53 \\
        &Jaw C&-0.12&0.63 \\ \hline
        \multirow{3}{*}{Subject 2}&Temple A&-0.15&0.57 \\
        &Temple B&0.17&-0.55 \\
        &Cheek C&0.12&0.50 \\ \hline
        \multirow{2}{*}{Subject 3}&Cheek A&0.09&0.42 \\
        &Jaw B&0.06&0.43 \\ \hline
        \multirow{3}{*}{Subject 4}&Forehead A&-0.02&0.54 \\
        &Cheek B&-0.08&0.72 \\
        &Cheek C&-0.09&0.66 \\ \hline
        \multirow{3}{*}{Subject 5}&Inner corner A&0.10&0.62 \\
        &Nose B&-0.06&0.57 \\
        &Outer corner C&0.12&0.59 \\ \hline
        \multirow{3}{*}{Subject 6}&Cheek A&0.10&0.61 \\
        &Cheek B&-0.10&0.59 \\
        &Jaw C&-0.09&0.60 \\ \hline
        \multirow{3}{*}{Subject 7}&Temple A&-0.13&0.55 \\
        &Cheek B&-0.07&0.57 \\
        &Cheek C&-0.11&0.57 \\ \hline
        \multirow{3}{*}{Subject 8}&Temple A&0.09&-0.63 \\
        &Neck B&0.12&-0.69 \\
        &Jaw C&0.13&0.59 \\
        \bottomrule
    \end{tabular}
\end{table}

\begin{figure}[tbp]
    \centering
    \begin{minipage}{.47\linewidth}
        \centering
        \subcaptionlistentry{Raw thermal image}
        \label{fig:raw5}
        \begin{overpic}[width=.7\linewidth]{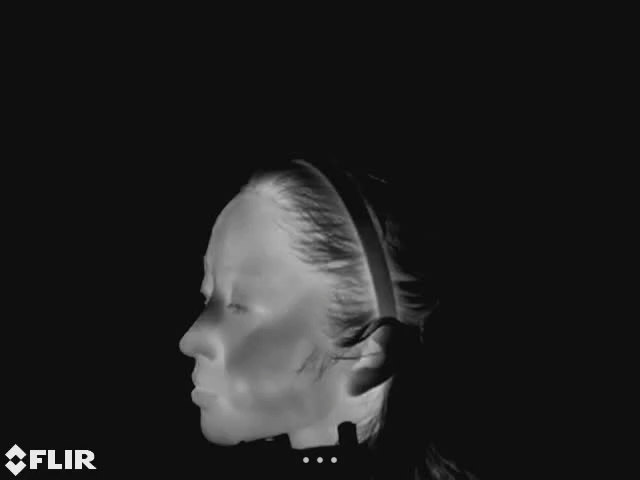}
            \put(-20,5){\colorbox{white}{\captiontext*{}}}
        \end{overpic}
    \end{minipage}
    \begin{minipage}{.47\linewidth}
        \centering
        \subcaptionlistentry{Proposed thermal image}
        \label{fig:synchro5}
        \begin{overpic}[width=.7\linewidth]{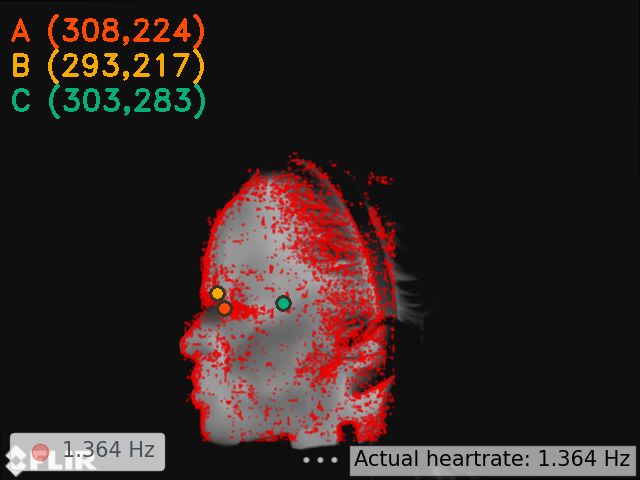}
            \put(-20,5){\colorbox{white}{\captiontext*{}}}
        \end{overpic}
    \end{minipage}
    \begin{minipage}{.47\linewidth}
        \subcaptionlistentry{Raw heartbeats}
        \label{fig:rawECG5}
        \begin{overpic}[width=\linewidth]{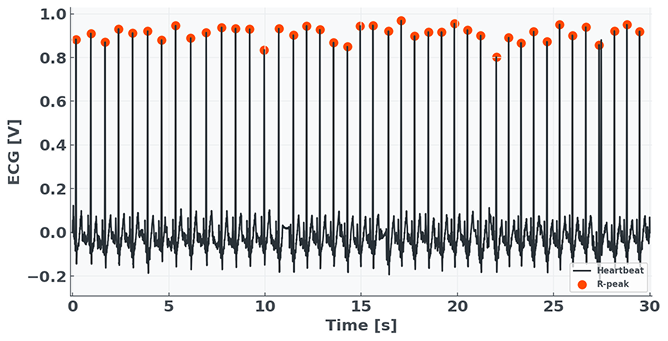}
            \put(0,5){\colorbox{white}{\captiontext*{}}}
        \end{overpic}
    \end{minipage}
    \begin{minipage}{.47\linewidth}
        \subcaptionlistentry{Normalized heartbeats}
        \label{fig:synchroECG5}
        \begin{overpic}[width=\linewidth]{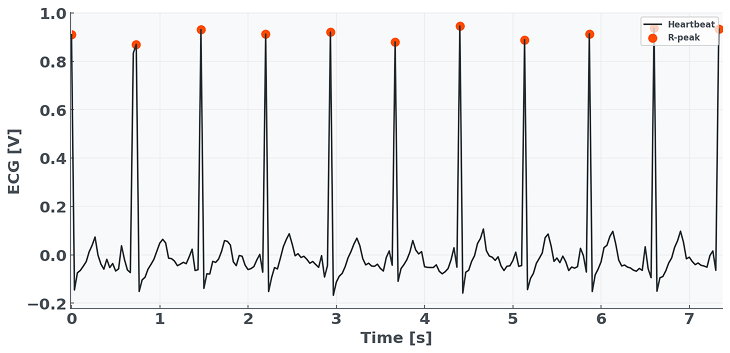}
            \put(0,5){\colorbox{white}{\captiontext*{}}}
        \end{overpic}
    \end{minipage}
    \begin{minipage}{.47\linewidth}
        \subcaptionlistentry{Raw temperature A}
        \label{fig:rawA5}
        \begin{overpic}[width=\linewidth]{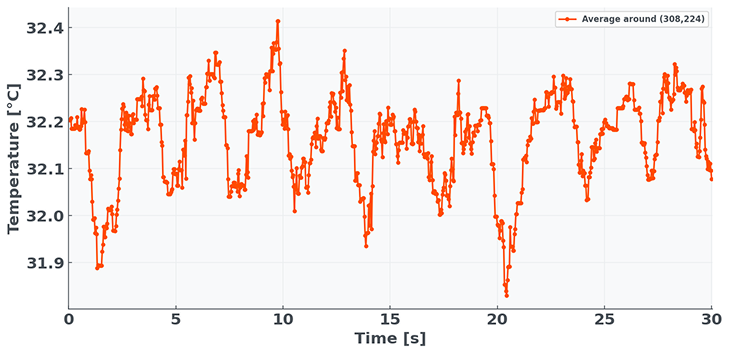}
            \put(0,5){\colorbox{white}{\captiontext*{}}}
        \end{overpic}
    \end{minipage}
    \begin{minipage}{.47\linewidth}
        \subcaptionlistentry{Proposed temperature A}
        \label{fig:synchroA5}
        \begin{overpic}[width=\linewidth]{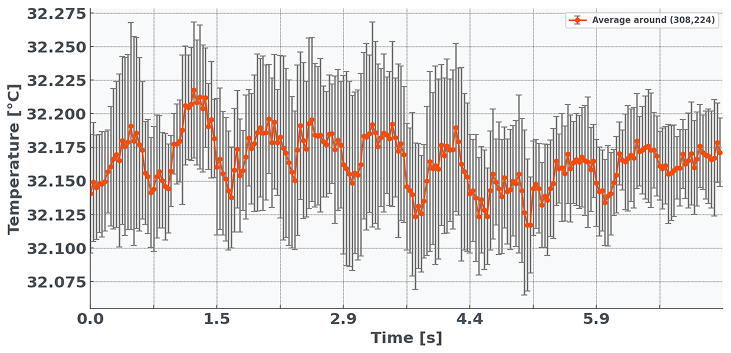}
            \put(0,5){\colorbox{white}{\captiontext*{}}}
        \end{overpic}
    \end{minipage}
    \begin{minipage}{.47\linewidth}
        \subcaptionlistentry{Raw temperature B}
        \label{fig:rawB5}
        \begin{overpic}[width=\linewidth]{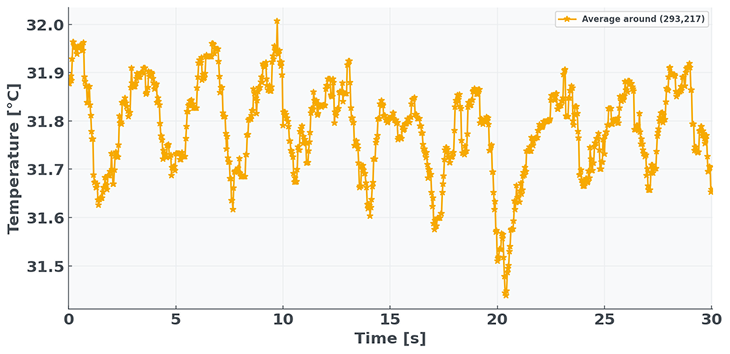}
            \put(0,5){\colorbox{white}{\captiontext*{}}}
        \end{overpic}
    \end{minipage}
    \begin{minipage}{.47\linewidth}
        \subcaptionlistentry{Proposed temperature B}
        \label{fig:synchroB5}
        \begin{overpic}[width=\linewidth]{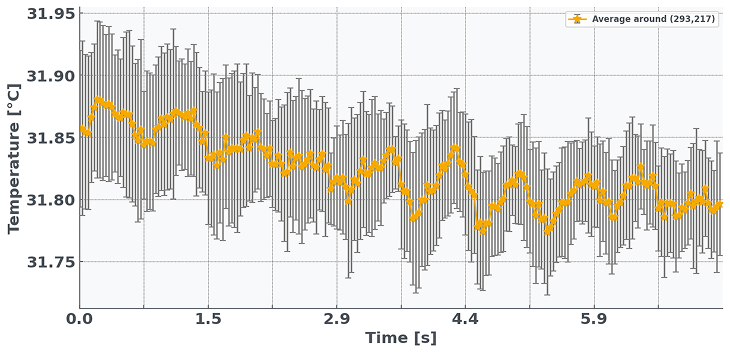}
            \put(0,5){\colorbox{white}{\captiontext*{}}}
        \end{overpic}
    \end{minipage}
    \caption{Raw data in left column and results of the proposed method in right column: Left profile of female subject 5. \subref{fig:raw5}, \subref{fig:synchro5} The thermal video frame. \subref{fig:rawECG5} Heartbeats of raw data. \subref{fig:synchroECG5} Heartbeats resampled to the mean RR interval. \subref{fig:rawA5}, \subref{fig:synchroA5} Temperature at the inner corner of the left eye A (308, 224). \subref{fig:rawB5}, \subref{fig:synchroB5} Temperature at the nose B (293, 217).} 
    \label{fig:5subject}
\end{figure}

\begin{figure}[tbp]
    \centering
    \begin{minipage}{.47\linewidth}
        \centering
        \subcaptionlistentry{Raw thermal image}
        \label{fig:raw6}
        \begin{overpic}[width=.7\linewidth]{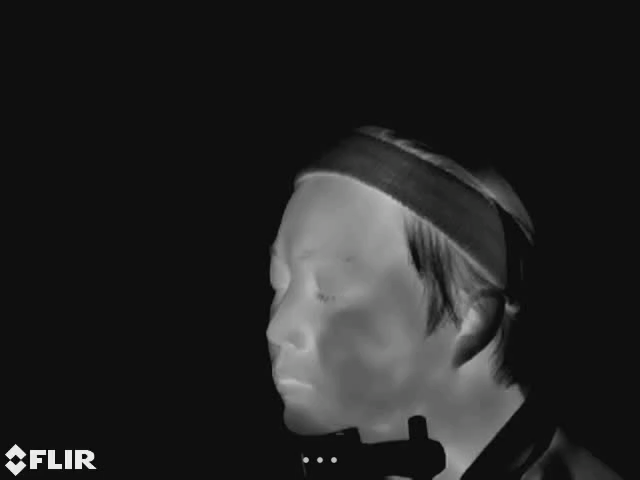}
            \put(-20,5){\colorbox{white}{\captiontext*{}}}
        \end{overpic}
    \end{minipage}
    \begin{minipage}{.47\linewidth}
        \centering
        \subcaptionlistentry{Proposed thermal image}
        \label{fig:synchro6}
        \begin{overpic}[width=.7\linewidth]{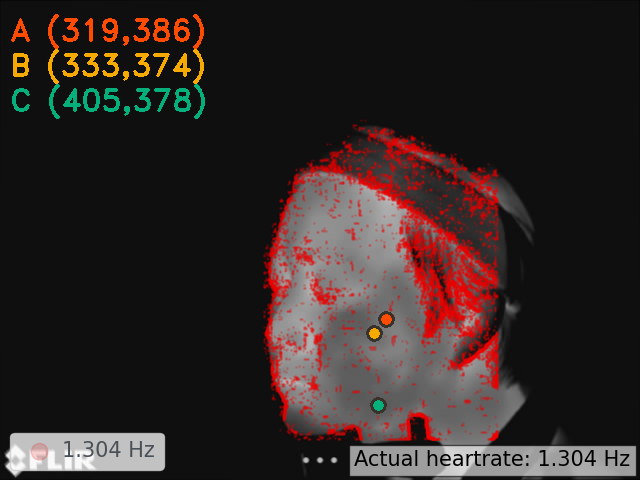}
            \put(-20,5){\colorbox{white}{\captiontext*{}}}
        \end{overpic}
    \end{minipage}
    \begin{minipage}{.47\linewidth}
        \subcaptionlistentry{Raw heartbeats}
        \label{fig:rawECG6}
        \begin{overpic}[width=\linewidth]{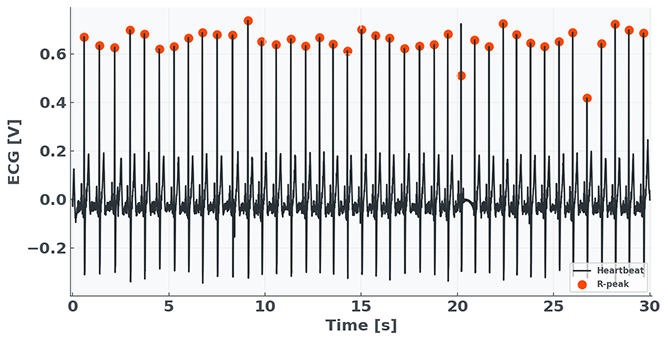}
            \put(0,5){\colorbox{white}{\captiontext*{}}}
        \end{overpic}
    \end{minipage}
    \begin{minipage}{.47\linewidth}
        \subcaptionlistentry{Normalized heartbeats}
        \label{fig:synchroECG6}
        \begin{overpic}[width=\linewidth]{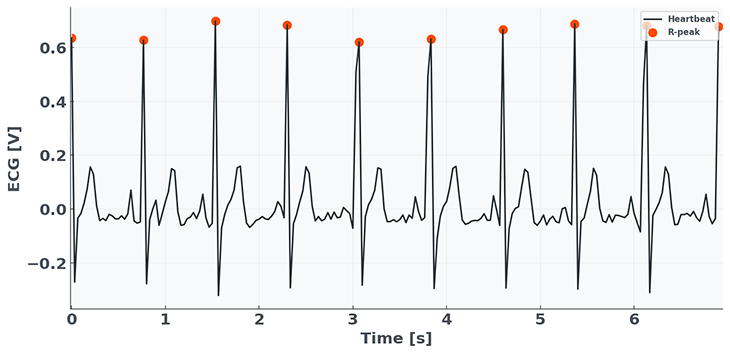}
            \put(0,5){\colorbox{white}{\captiontext*{}}}
        \end{overpic}
    \end{minipage}
    \begin{minipage}{.47\linewidth}
        \subcaptionlistentry{Raw temperature A}
        \label{fig:rawA6}
        \begin{overpic}[width=\linewidth]{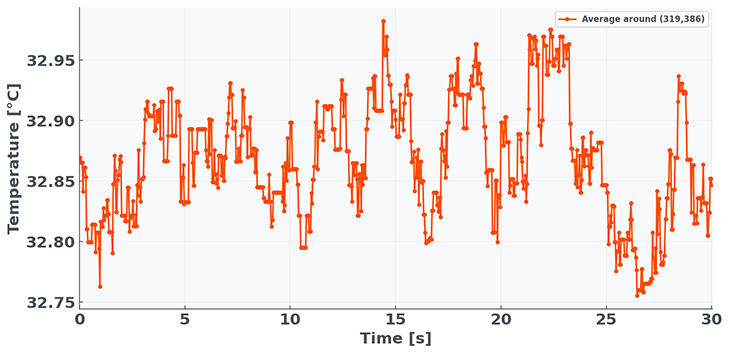}
            \put(0,5){\colorbox{white}{\captiontext*{}}}
        \end{overpic}
    \end{minipage}
    \begin{minipage}{.47\linewidth}
        \subcaptionlistentry{Proposed temperature A}
        \label{fig:synchroA6}
        \begin{overpic}[width=\linewidth]{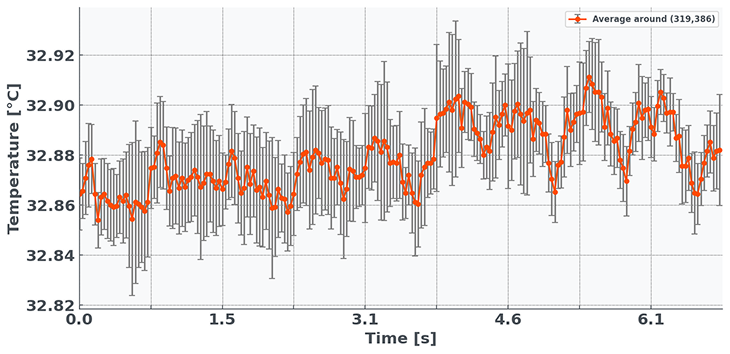}
            \put(0,5){\colorbox{white}{\captiontext*{}}}
        \end{overpic}
    \end{minipage}
    \begin{minipage}{.47\linewidth}
        \subcaptionlistentry{Raw temperature B}
        \label{fig:rawB6}
        \begin{overpic}[width=\linewidth]{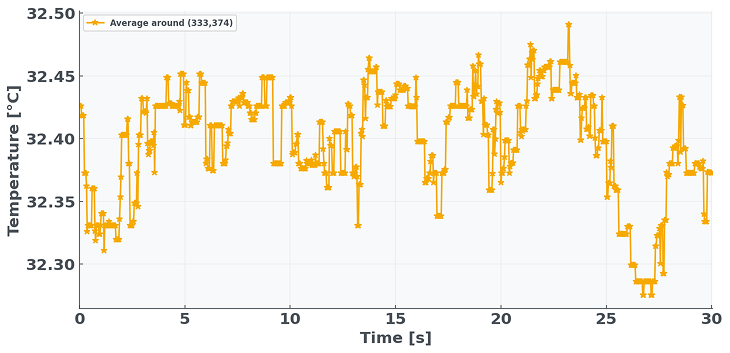}
            \put(0,5){\colorbox{white}{\captiontext*{}}}
        \end{overpic}
    \end{minipage}
    \begin{minipage}{.47\linewidth}
        \subcaptionlistentry{Proposed temperature B}
        \label{fig:synchroB6}
        \begin{overpic}[width=\linewidth]{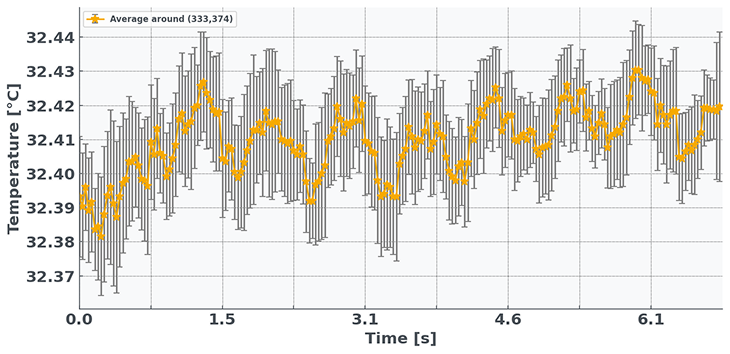}
            \put(0,5){\colorbox{white}{\captiontext*{}}}
        \end{overpic}
    \end{minipage}
    \caption{Raw data in left column and results of the proposed method in right column: Left profile of male subject 6. \subref{fig:raw6}, \subref{fig:synchro6} The thermal video frame. \subref{fig:rawECG6} Heartbeats of raw data. \subref{fig:synchroECG6} Heartbeats resampled to the mean RR interval. \subref{fig:rawA6}, \subref{fig:synchroA6} Temperature at the left cheek A (319, 386). \subref{fig:rawB6}, \subref{fig:synchroB6} Temperature at the left cheek B (333, 374).} 
    \label{fig:6subject}
\end{figure}

\begin{figure}[tbp]
    \centering
    \begin{minipage}{.47\linewidth}
        \centering
        \subcaptionlistentry{Raw thermal image}
        \label{fig:raw7}
        \begin{overpic}[width=.7\linewidth]{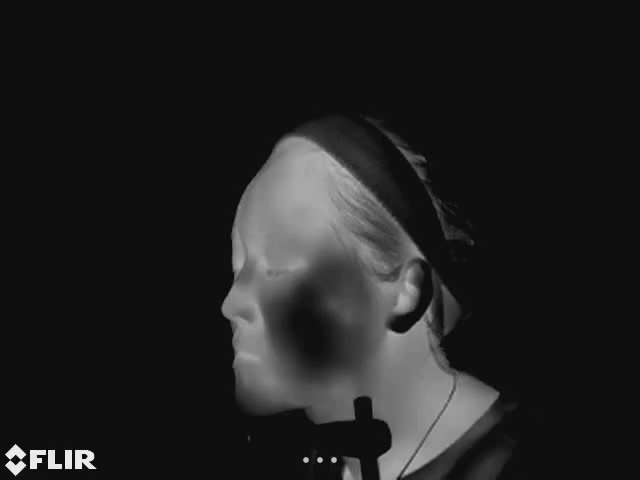}
            \put(-20,5){\colorbox{white}{\captiontext*{}}}
        \end{overpic}
    \end{minipage}
    \begin{minipage}{.47\linewidth}
        \centering
        \subcaptionlistentry{Proposed thermal image}
        \label{fig:synchro7}
        \begin{overpic}[width=.7\linewidth]{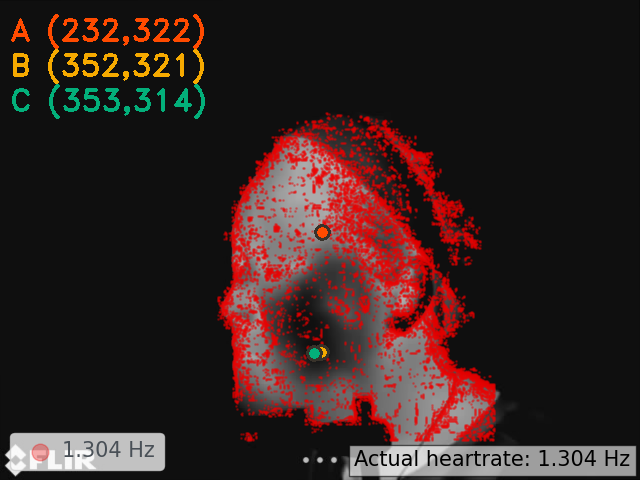}
            \put(-20,5){\colorbox{white}{\captiontext*{}}}
        \end{overpic}
    \end{minipage}
    \begin{minipage}{.47\linewidth}
        \subcaptionlistentry{Raw heartbeats}
        \label{fig:rawECG7}
        \begin{overpic}[width=\linewidth]{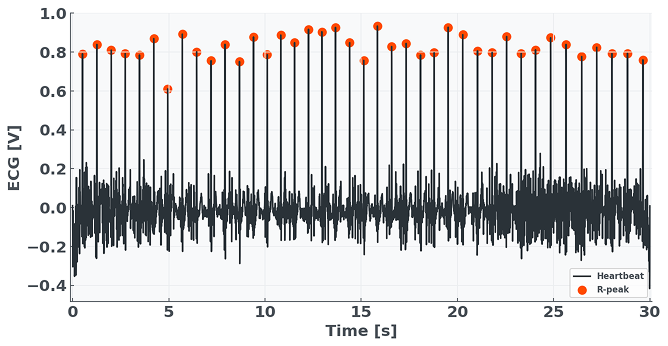}
            \put(0,5){\colorbox{white}{\captiontext*{}}}
        \end{overpic}
    \end{minipage}
    \begin{minipage}{.47\linewidth}
        \subcaptionlistentry{Normalized heartbeats}
        \label{fig:synchroECG7}
        \begin{overpic}[width=\linewidth]{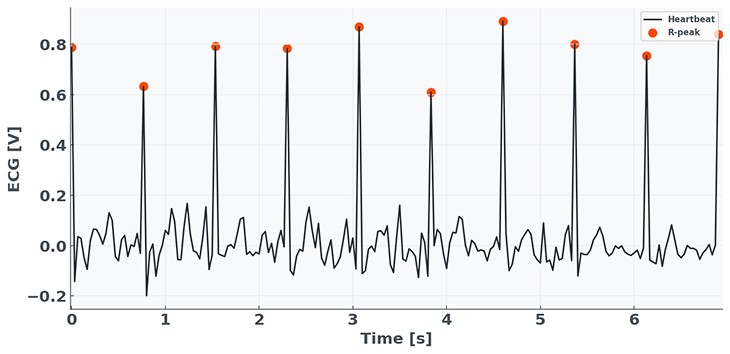}
            \put(0,5){\colorbox{white}{\captiontext*{}}}
        \end{overpic}
    \end{minipage}
    \begin{minipage}{.47\linewidth}
        \subcaptionlistentry{Raw temperature A}
        \label{fig:rawA7}
        \begin{overpic}[width=\linewidth]{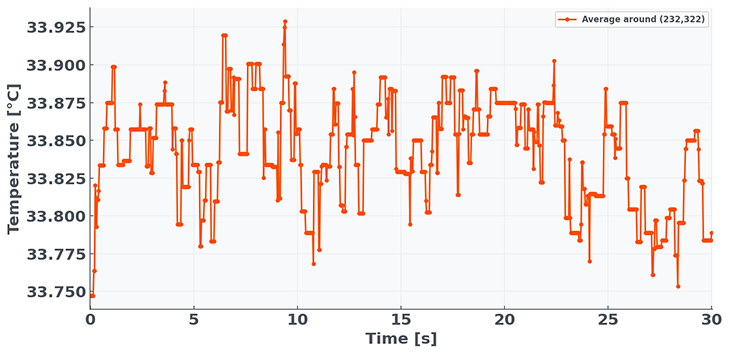}
            \put(0,5){\colorbox{white}{\captiontext*{}}}
        \end{overpic}
    \end{minipage}
    \begin{minipage}{.47\linewidth}
        \subcaptionlistentry{Proposed temperature A}
        \label{fig:synchroA7}
        \begin{overpic}[width=\linewidth]{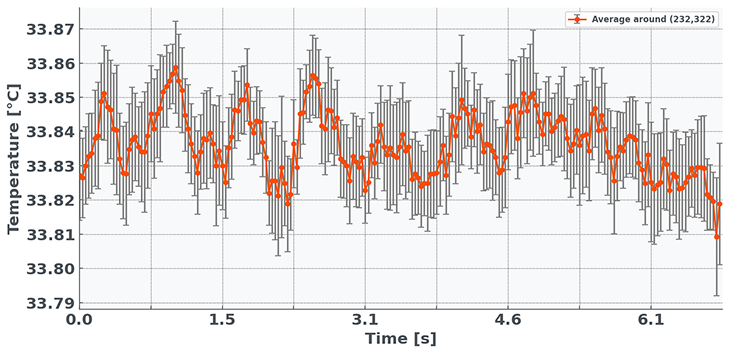}
            \put(0,5){\colorbox{white}{\captiontext*{}}}
        \end{overpic}
    \end{minipage}
    % \begin{minipage}{.47\linewidth}
    %     \begin{overpic}[width=\linewidth]{fig/7subject/24_1319_raw1.png}
    %     \subcaptionlistentry{Raw temperature B}
    %     \label{fig:rawB7}
    % \end{minipage}
    % \begin{minipage}{.47\linewidth}
    %     \begin{overpic}[width=\linewidth]{fig/7subject/24_1319_err1.png}
    %     \subcaptionlistentry{Proposed temperature B}
    %     \label{fig:synchroB7}
    % \end{minipage}
    \begin{minipage}{.47\linewidth}
        \subcaptionlistentry{Raw temperature C}
        \label{fig:rawC7}
        \begin{overpic}[width=\linewidth]{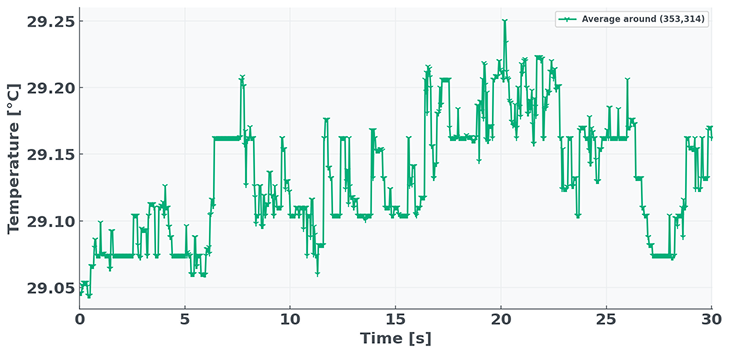}
            \put(0,5){\colorbox{white}{\captiontext*{}}}
        \end{overpic}
    \end{minipage}
    \begin{minipage}{.47\linewidth}
        \subcaptionlistentry{Proposed temperature C}
        \label{fig:synchroC7}
        \begin{overpic}[width=\linewidth]{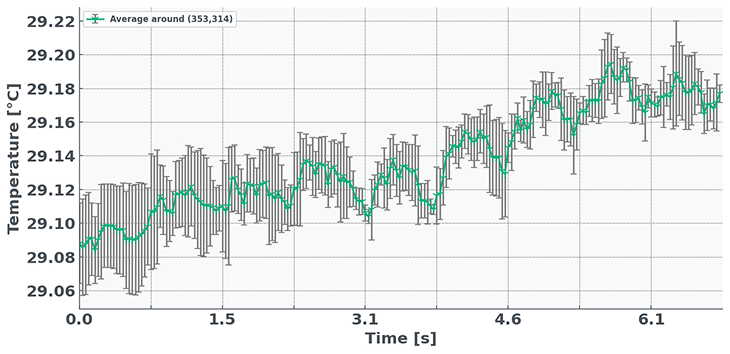}
            \put(0,5){\colorbox{white}{\captiontext*{}}}
        \end{overpic}
    \end{minipage}
    \caption{Raw data in left column and results of the proposed method in right column: Left profile of female subject 7. \subref{fig:raw7}, \subref{fig:synchro7} The thermal video frame. \subref{fig:rawECG7} Heartbeats of raw data. \subref{fig:synchroECG7} Heartbeats resampled to the mean RR interval. \subref{fig:rawA7}, \subref{fig:synchroA7} Temperature at the left temple A (232, 322). \subref{fig:rawC7}, \subref{fig:synchroC7} Temperature at the left cheek C (353, 314).} 
    \label{fig:7subject}
\end{figure}

Specific results for body temperature and heart rate are then presented in \cref{fig:5subject,fig:6subject,fig:7subject}.
The results for female subject 5 are presented in \cref{fig:5subject}. The left column of \cref{fig:5subject} shows the raw data, and the right column shows the results after applying synchro-thermography. The error bars in \cref{fig:synchroA5,fig:synchroB5} represent the standard errors shown in \cref{eq:standardError}.
Regarding the results of subject 5, we speculate that the inner corner of the left eye A was affected by the medial palpebral arteries, the nose B by the dorsal nasal artery, and the outer corner of the left eye C by the vascular activity of the zygomatico-orbital artery.

Next, the results are shown for male subject 6 in \cref{fig:6subject}. For subject 6, the left cheeks A and B are likely affected by the infra-orbital artery and the transverse facial artery, and the jaw C is likely affected by the facial artery.

The results for female subject 7 are shown in \cref{fig:7subject}.
In the results for subject 7, the left temple A is due to the frontal branch of the superficial temporal artery, while the left cheek B and C show temperature fluctuations due to the infra-orbital artery and facial artery vessels. It is possible that the temperature fluctuations are represented in \cref{fig:synchroA7,fig:synchroC7}.

Consequently, regardless of gender, body temperature changes accompanying heart rate variability could be measured at the foreheads, temples, cheeks, noses, and necks. The results of the preliminary experiment and the main experiment indicate that the proposed method can be applied regardless of the orientation of the face, since the body temperature changes accompanying heart rate variability could be measured not only in the frontal face but also in the profile face.

\subsection{Discussion}
From the experiment, it was confirmed that changes in body temperature associated with heart rate activity were observed in all the subjects using the proposed method. However, since the number of subjects in this study was limited and all subjects were in their 20s and of Asian ethnicity (Japanese and Chinese), further validation with a broader range of age groups and ethnicities is needed.

In some periods in \cref{fig:synchroB6,fig:synchroC7}, the waveforms are unclear or have small amplitudes in some parts. One possible reason for this is that the noise is not completely removed by the proposed method. This could be improved by increasing the window size $T_{win}$, which results in the measurement time being longer than 10 heartbeats. Alternatively, the signal-to-noise ratio could be further improved by averaging the regions where the waveforms of the temperature fluctuations are in-phase in the spatial direction as well. For example, in subject 7's temple A (\cref{fig:synchroA7}) and cheek C (\cref{fig:synchroC7}), the periods of temperature fluctuation are close to in-phase. We believe that additive averaging in these regions can improve the signal-to-noise ratio while keeping the measurement time short.

The physiological basis of our findings is not entirely clear at the moment. We presume that changes in facial skin temperature measured using synchro-thermography may reflect vascular activity, such as in arteries and arteriovenous anastomoses (AVA) \cite{Chen2019BEL,Jose2022FP}. The fact that skin surface temperatures changed with the same periodicity as the heartbeat interval supports the concept that body temperature homeostasis is maintained by blood circulation. Further multifaceted analysis using other physiological indices such as blood flow, pulse transit time, and blood pressure should be conducted in the future.

\section{Conclusion}
We developed Synchro-Thermography, a novel methodology to measure body temperature changes associated with blood flow activity for physiological sensing. The proposed method utilizes the heartbeat as a reference signal for timing, allowing for the measurement of minute changes in body temperature without the need for external stimuli. Therefore, the proposed method can be applied to the head and neck region, which is a difficult area to stimulate externally. Moreover, the proposed method can be used in a wide range of applications because there is no extra stimulus or task for the subject to perform during the measurement, so the natural response of the subject can be measured.

The proposed method can improve the signal-to-noise ratio by a factor of 2 or more when the noise follows a normal distribution by taking a synchronous additive average of 10 heartbeats (about 8 seconds). Experiments were conducted with eight subjects and small skin surface temperature changes of about \SI{10}{mK} in amplitude were measured for both males and females. The time of synchronized additive averaging was calibrated to each subject by matching the time of synchronized additive averaging to the heartbeat interval of each individual. In this experiment, electrodes were worn and electrocardiograms were measured to ensure that the measurements were accurate. However, the proposed method can be applied in principle to pulses obtained from wearable devices such as smartwatches as well. In this experiment, a webcam was also used to record visible video for the entire experiment. Therefore, it is also possible to perform the measurements completely remotely by estimating the pulse rate from the RGB video \cite{Hsu2017IJCB,Mehta2023TIM}. Synchro-thermography can also be applied using a method of non-contact measurement of the heartbeat \cite{Matsumoto2020NE}.

In the future, it is necessary to clarify the physiological basis of our research results using other physiological indices such as blood flow and blood pressure. At the same time, we will investigate how to apply the skin surface temperature changes associated with blood flow activity measured by the proposed method to physiological sensing, such as identifying areas where blood flow changes have occurred due to injury or disease, and measuring stress. Since the proposed method is easy to set up, it can be used not only in the laboratory but also for sensing dynamic physiological information that occurs in real life, such as measuring emotions when eating or drinking.

% Do not put page numbers in the manuscript PDF.

\section*{Acknowledgements}
This work was supported by JSPS Grant-in-Aid for Scientific Research (A) Grant Number JP24H00704 and Suntory Holdings Limited.

\bibliographystyle{ieeetr}
\bibliography{ref}

\end{document}